\documentclass[prl,twocolumn,preprintnumbers,amsmath,amssymb]{revtex4}

\usepackage{graphicx}% Include figure files
\usepackage{bm}%bold greek letters
\usepackage{color} 
\usepackage{amssymb, dsfont}

\newcommand{\BEQ}{\begin{equation}}
\newcommand{\EEQ}{\end{equation}}

\newcommand{\beq}{\begin{equation}}
\newcommand{\eeq}{\end{equation}}
\newcommand{\bea}{\begin{eqnarray}}
\newcommand{\eea}{\end{eqnarray}}

\begin{document}

 \begin{@twocolumnfalse}
 \begin{center}
    \mbox{
{
\color{blue} \href{http://dx.doi.org/10.1103/PhysRevLett.115.188303}{DOI: 10.1103/PhysRevLett.115.188303}
}
    }
\end{center}
  \end{@twocolumnfalse}

\title{Self-Sustained Density Oscillations of Swimming Bacteria Confined in Microchambers}
%\title {Spontaneous Flows and Accumulation of Active Particles in Connected Micro-Chambers}

\author{M.~Paoluzzi$^{1}$}
\email{matteo.paoluzzi@ipcf.cnr.it}
\author{R.~Di Leonardo$^{1,2}$}
\author{L.~Angelani$^{1,3}$}
\affiliation{
$^1$ Dipartimento di Fisica Universit\`a Sapienza, P.le A Moro 2, 00185 Rome, Italy \\ 
$^2$ NANOTEC-CNR, Institute of Nanotechnology, Soft and Living Matter
Laboratory, Piazzale A. Moro 2, I-00185, Roma, Italy \\ 
$^3$ Istituto dei Sistemi Complessi (ISC-CNR), UOS Sapienza, P.le A Moro 2, 00185 Rome, Italy}

\begin{abstract}
We numerically study the dynamics of run-and-tumble particles confined in two chambers connected by
thin channels.  Two dominant dynamical behaviors emerge: ({\it i}) an oscillatory pumping state, in which
particles periodically fill the two vessels and ({\it ii}) a circulating flow state, dynamically maintaining
a near constant population level in the containers when connected by two channels.
We demonstrate that the oscillatory behaviour arises from the combination of a narrow channel, preventing bacteria reorientation, and a density dependent motility inside the chambers.
\end{abstract}

\maketitle

%%%%%%%%%%%%%%%  TEXT  %%%%%%%%%%%%%%%%%%%%%%%%%%%% 
{\it Introduction.}--- 
Self-sustained oscillators are ubiquitous in physics and biology \cite{sync}.  
From the Van der Pol oscillator to the heartbeat these systems are characterised by a 
periodic motion that is sustained against dissipation by some form of energy source.
Active systems constantly consume an internal energy source to sustain persistent motions 
in a highly dissipative environment \cite{Berg04,Marchetti13}. 
This results in a strongly out of equilibrium dynamics
that can give rise to 
unidirectional actuation of micro-objects \cite{Angelani09,Angelani11c,DiLeonardo10,Sokolov10},
spontaneous accumulation of passive colloids over target sites \cite{Koumakis13},
long lived density fluctuations \cite{Narayan07,Toner,Ramaswamy03},
large-scale vortex lattice \cite{Sumino12}, frozen steady states \cite{Schaller11}, 
active liquid crystals \cite{Zhou14} or macroscopic directed motion \cite{Bricard13}.
It has been shown that the large scale collective patterns that spontaneously break and reform in many active systems can be stabilized by confining microstructures \cite{Sanchez12, wioland, Galajda14}.
However spatial confinement always results in long lived unidirectional flows that only rarely and randomly can switch direction \cite{Galajda14}.

In this Letter we investigate the possibility of using geometric confinement to obtain self-sustained oscillations in active matter systems. In particular, by using numerical simulations, we show that active particles
can alternately fill and empty two micro chambers connected by thin channels.
We find that the number of particles inside each chamber fluctuates with a distribution that becomes increasingly bivariate for large particle densities. 
The narrowness of the channels ensures single file dynamics \cite{Jelic12,Illien13,Gorissen12,Ner13,Locatelli14,Wei00} 
and inhibits cell reorientation. 
This, together with a density dependent motility of swimmers inside the chambers, gives rise to the observed oscillatory behavior.
In addition, when the chambers are connected by two channels we observe also a circulating flow maintaining a near constant population level in the containers.
%%%%%%%%%%%%%%%%%%%%%%%%%%%%%%%%%%%%%%%%%%%%%%%%%%%%%%%%%%%5

\noindent
{\it Results and discussion.}---
We perform Molecular Dynamics simulations of $N$ run-and-tumble
swimmers \cite{Schnitzer93, Tailleur08, Cates12, Reichhardt08}  of length $\ell$ and thickness $a$, with $a/\ell=1/2$, in two dimensions \cite{Angelani09,Angelani11b,Paoluzzi}.
The interaction between the swimmers is purely repulsive.
The system is confined in two chambers connected by one or two channels. Details of the model can be found in \cite{supp}.

\begin{figure}[!t]
\centering
\includegraphics[width=.47\textwidth]{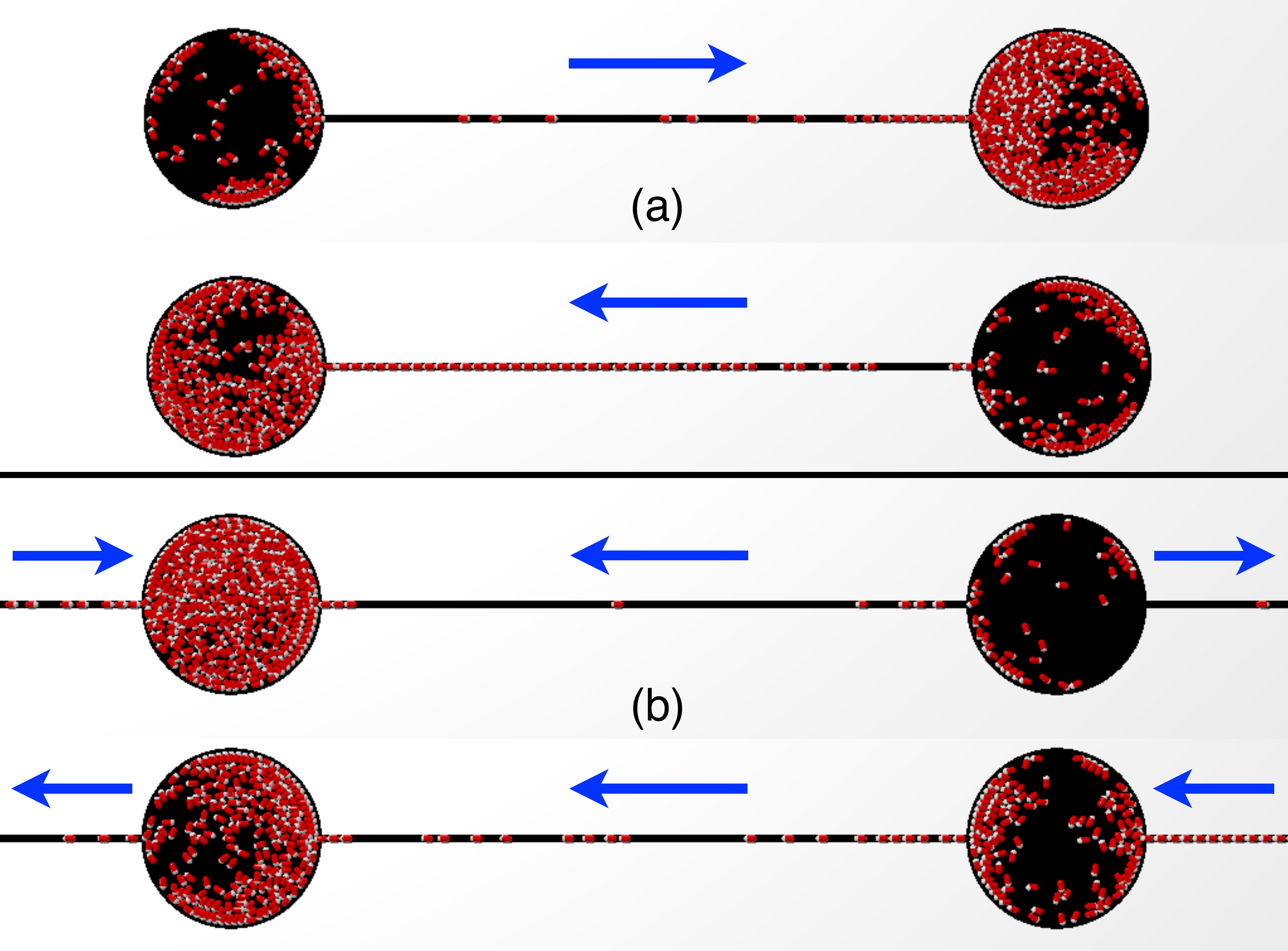}
\caption{ 
Snapshots of the simulations of active particles confined in two chambers connected by thin channels.
Panel (a): alternating pumping, in which particles  alternately fill and empty the chambers in the case of a single channel.
Panel (b): alternating pumping and circular flow in the case of two channels connecting the chambers.}
\label{fig:snap}       % Give a unique label
\end{figure}

\begin{figure}[!t]
\centering
\includegraphics[width=.47\textwidth]{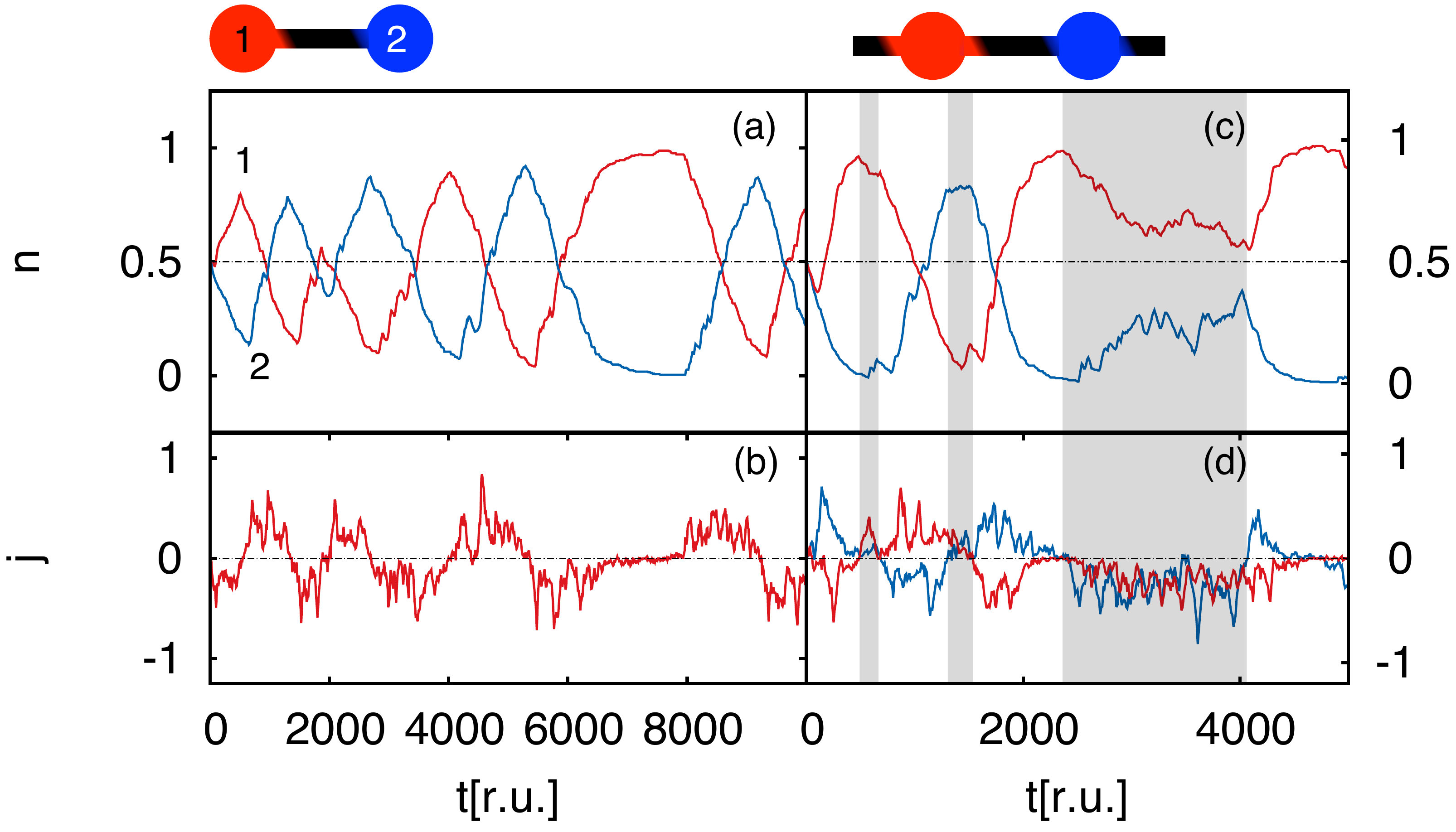}
\caption{
Time evolution of the fraction of swimmers $n_{1,2}(t)$ inside the vessels and the channel current $j(t)$
for a typical single run.
Left panels refer to the case of one channel ($\phi\!=\!0.47$),
while right panels to the case of two channels ($\phi\!=\!0.44$).
In panel (d) the currents of the first and second channel are reported
and grey area correspond to concordant currents, $j_1 \cdot j_2 >0$.
} 
\label{fig:circular}       % Give a unique label
\end{figure}

\begin{figure}[!t]
\centering
\includegraphics[width=.5\textwidth]{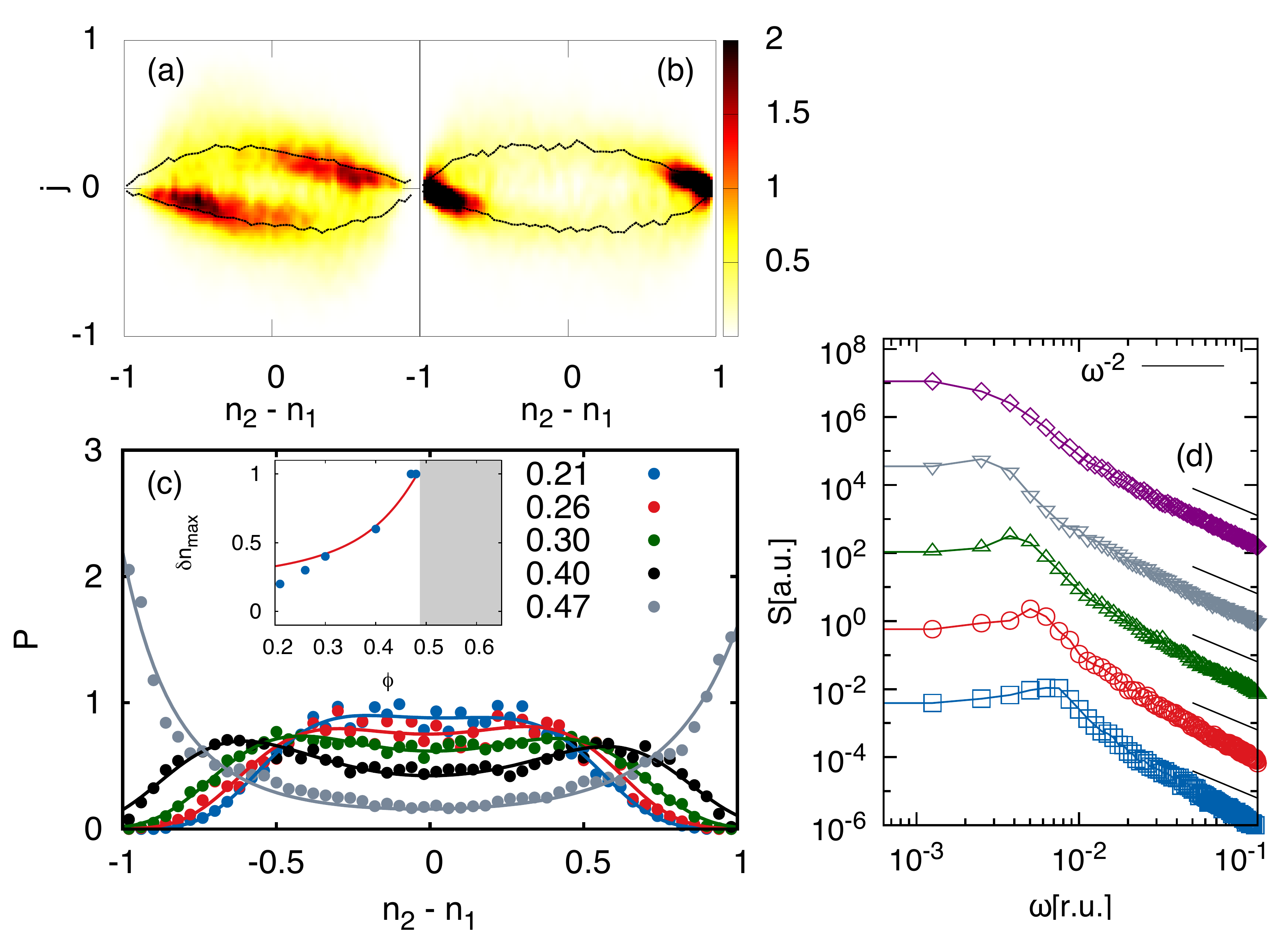}
\caption{
Single channel.
Top panels: the join probability $P(n_2\!-\!n_1,j)$ at two different densities, $\phi=0.30$ (panel (a))
and $\phi\!=\!0.47$ (panel (b)).
Panel (c): probability distribution function
$P(n_2\!-\!n_1)$ increasing the number of swimmers
for $L=50\ell$ and $\sigma=\ell/2$. 
The full lines are fits with $f(\delta n)=\exp(a\!+\!b \delta n^2\!+\!c \delta n^4)$.
Inset: %variance of $P(\delta n)$ (red symbols) and
$\delta n_{max}$ (blue symbols) as a function of $\phi$, the red line is the amplitude of the oscillations predicted by the model.
Densities in the grey area correspond to jammed states where all bacteria fill one of the two chambers almost completely.
Panel (d):
power spectrum $S(\omega)$ at five different densities.
For clarity the individual curves have been vertically shifted 
multiplying for $10^{2n}$ with $n=0,...,5$. The
corresponding area fractions $\phi$ are (from bottom to top) 
$0.21$ (blue symbols), $0.26$ (red),
$0.30$ (green), $0.40$ (grey) and $0.47$ (purple).
}
\label{fig:rho2}       % Give a unique label
\end{figure}

We first consider two circular chambers of radius $R\!=\!7\ell$ connected by a thin channel of length $L\!=\!50\ell$ and transverse size $\sigma\!=\!\ell/2$ (see Fig. (\ref{fig:snap}), panel (a)).   
The total number of particles $N$ was varied from 160 to 464 corresponding to area fractions $\phi$ going from 0.21 to 0.62 \cite{phi}.
We found that 
the presence of persistent currents in thin channels
gives rise to large fluctuations in particle distributions between the two chambers.
The symmetric state, where the two chambers are equally populated, becomes unstable due to the spontaneous formation of large currents with a characteristic lifetime that increases with the total number of swimmers.  As a consequence, one of the two chambers is progressively filled with particles up to a point where it triggers the formation of a reversed current. This mechanism gives rise to an oscillatory pumping between the two containers as evidenced in the snapshots of the numerical simulations (panel (a), Fig. (\ref{fig:snap}) and the movie in \cite{supp}).
In Fig. (\ref{fig:circular}a) the time evolution of the fraction of particles in the two reservoirs $n_{1,2}$ are shown for a sample run with $\phi=0.47$.
Fig. (\ref{fig:circular}b) shows the corresponding channel current defined as the sum of the velocities of particles inside the channel, $j=\sum_k^\prime v_k(t)/L$. A clear oscillation is observed 
with a period of about $2000$. 
In the following we will consider the probability distribution for
coarse grained variables $P (x)$ and $P (x, y)$, defined as $P(x)\equiv \overline{\langle\delta(x - x(t))\rangle}$ and $P(x,y)\equiv \overline{\langle\delta(x - x(t)) \delta(y-y(t)) \rangle}$ 
where $\overline{\cdots}$ indicates the time average and the angular brackets $\langle \ldots \rangle$ averages over $120$ independent runs.
Calling $\delta n=n_2\!-\!n_1$ the asymmetry parameter, we compute the joint probability density  $P(\delta n,j)$ and  
the power spectrum $S(\omega)=|\delta \hat{n}(\omega)|^2$,
 with $\delta \hat{n}(\omega)$ the Fourier transform of $\delta n(t)$.
In the upper panels of Fig. (\ref{fig:rho2}) the quantity $P(\delta n,j)$ is shown for $\phi=0.30$ and $\phi=0.47$. Contrary to the case of passive thermal particles, where a stable equilibrium point exists  at ($\delta n\!=\!0$, $j\!=\!0$), here we observe a stationary limit cycle corresponding to fluxes that alternately empty and fill the containers  (a video of a typical trajectory followed by the system in the $(\delta n,j)$ plane is shown in Supplemental Material \cite{supp}.
%%%%%%%%%%%%%%%%%%%%%%%%%%%%%%%%
This oscillating behaviour appears as a peak in the power spectrum $S(\omega)$ (Fig. (\ref{fig:rho2})d).
The high frequency behavior of $S(\omega)$ is well approximated by a $\sim\omega^{-2}$ tail (the solid black lines in Fig. (\ref{fig:rho2})d), 
indicating that the dynamics of $\delta n(t)$ is uncorrelated on small time-scales.
When the number of particles is increased, it takes a longer time to fill or empty the chambers resulting in a decrease of the peak frequency.
For high particle densities $\delta n$ displays fluctuations that are large and long lived
as shown by the pronounced peaks at $\delta n\sim\pm 1$ in $P(\delta n,j)$ (Fig. 3b).  The two chambers alternately fill and empty almost completely although transitions between these two states lose periodicity and the peak in $S(\omega)$ disappears (Fig. 3d). 

In order to analyze in more details the role played by the swimmers density
we investigate the behavior of $P(\delta n)$ at different particles number.
As one can see in panel (c) of Fig. (\ref{fig:circular}), 
the distribution $P(\delta n)$ becomes bimodal by increasing $\phi$.
The bimodality of $P(\delta n)$ indicates that density oscillations
take place between two high density and long lived states, as a result the
two chambers are alternatively almost filled and $\delta n\to\pm 1$.
%The variance of the distribution and the quantity $\delta n_{max}$, defined
The quantity $\delta n_{max}$, defined as the value of $|\delta n|$ where
the distribution reaches its maximum value, continuously
grows withÊ $\phi$, up to the threshold value $\phiÊ = 0.47$.
For higher densities one of the two chambers fills up completely and
the corresponding internal motility vanishes \cite{supp} giving rise to long lived jammed states (the grey area in Fig. 3c).
%There is, then, a threshold density  
%above which the most probable states visited by the system are $\delta n=\pm1$. 
%\textcolor{blue}{When a chamber is filled above the threshold density, the resulting
%jammed phase is long lived and the system remain blocked in the phase $\delta n=\pm 1$ (see grey area in Fig. 3c).}
Varying the channel length $L$ from $50\ell$ to $25\ell$, at
least in the analyzed range, seems to have little effects on the results \cite{supp}. %\cite{supp}.
Performing simulations at different channel size $\sigma/\ell=1/2,1,3/2,2$
(with $L/\ell=25,50$ and for the case $\phi=0.47$, i.e. where $P(\delta n)$ is strongly bimodal) 
we observe that $\delta n_{max}$
decreases toward $0$ by increasing $\sigma$, indicating that the channel thickness 
plays a crucial role in the observed alternate pumping phenomenon \cite{supp}.

\begin{figure}[!t]
\centering
\includegraphics[width=.5\textwidth]{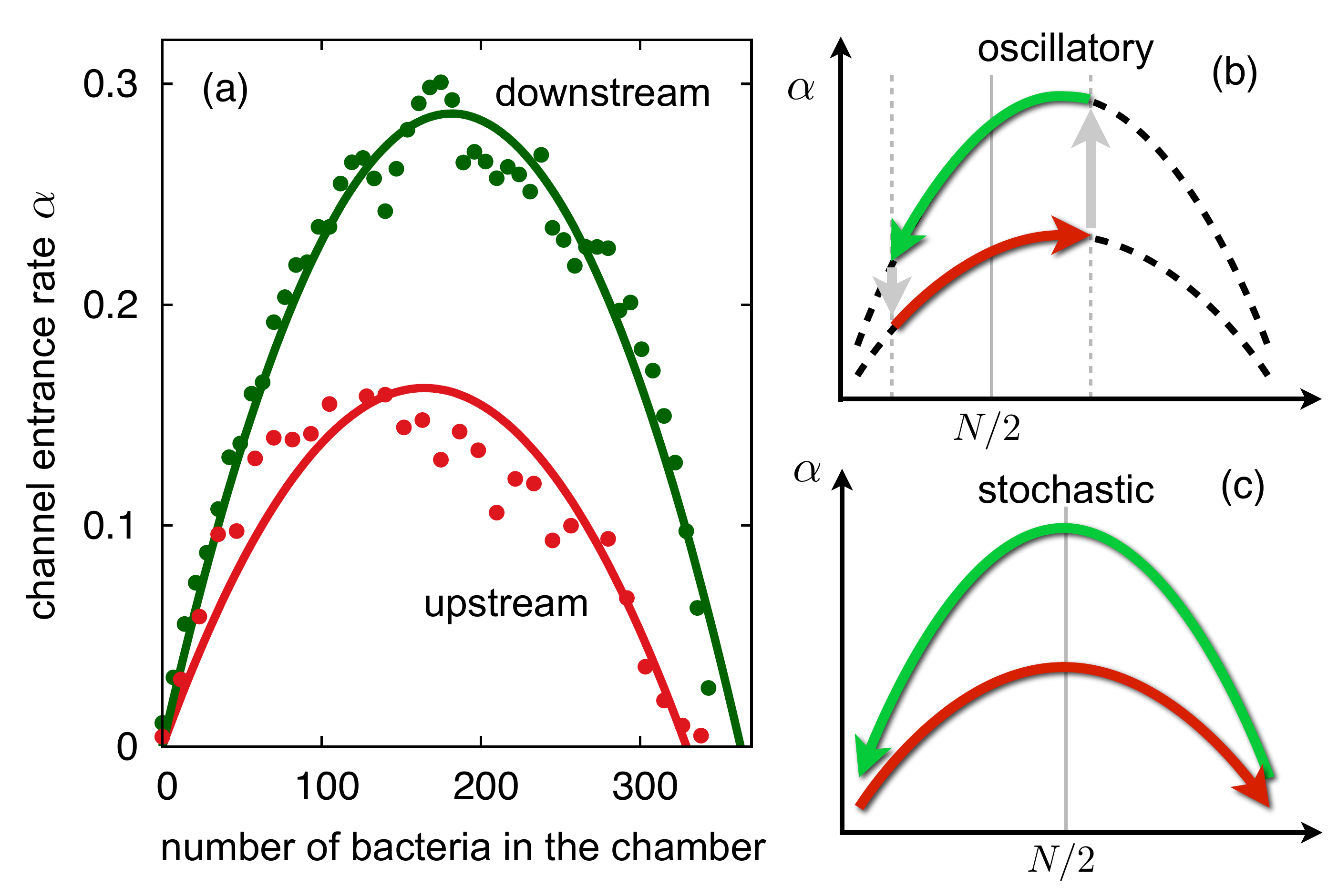}
\caption{
%\textbf{
Left panel: variation of the number of swimmers in the channel as a function of $N_i$ for downstream (green symbols)
and upstream (red symbols). The full lines are the fits with $\alpha$.
Right panels: the sketch of the mechanism responsible of the alternating pumping
for the oscillatory regime (top) and in the case of the stochastic regime (bottom),
i. e., when the current is reversed with no defined period. 
%}
}
\label{fig:fig5}       % Give a unique label
\end{figure}

At the microscopic level the mechanism responsible for the
observed phenomenology relies on the fact that particles in
a thin channel are unable to reverse their direction of motion,
thus forming long files of pushing bacteria.
A small difference in the number of left and right oriented bacteria
determines the sign of the net flux, and 
counter-current swimming bacteria are
then pushed back to their original chamber, with a consequent increase
of the net flux in the channel.
As a result small fluctuations are amplified,
giving rise to long lived flows of particles in the system.
For non interacting particles, the rate of bacteria passing from one chamber to the channel will be simply proportional to the number of bacteria in the chamber $N_i$ with $i=1,2$ the chamber index. 
Because of interactions, when the number of particles in the chamber is near to close packing $N_0$, the motility will be reduced with a consequent reduction in the entrance rate \cite{supp}. 
 A simple form for the rate that is consistent with this behaviour is  $\alpha=\beta N_i (1-N_i/N_0) $.
The rate of entrance in the channel is also affected by the sign of $j$ that makes it harder for bacteria to enter a channel when there's a counter-flowing current. To account for this last observation we suppose that  $\beta$ can assume two different values $\beta_+$, $\beta_-$ according to the sign of the current $j$. We found that the above expression for $\alpha$ fits very well our simulation data Fig. (\ref{fig:fig5}) panel (a).
The origin of the oscillating behaviour becomes now clear if one follows the time evolution of $N_{1,2}$ over the curves in Fig. (\ref{fig:fig5},a). Let's start from the case where the total number of particles $N$ is less than $N_0$ and bacteria are equally distributed in the two chambers. Due to the narrowness of the channel, bacteria will conserve their direction from the moment they enter the channel until they exit. A current reversal, for example from positive to negative, can then be triggered only if the rate of left going bacteria entering from the right chamber is higher than the number of right going bacteria entering from the left chamber. If we assume an initial positive current than the left chamber will loose bacteria and $N_1$ will move on the upper curve in Fig. (\ref{fig:fig5},b) towards left. Since the total number of particles has to be conserved $N_2$ will move symmetrically on the lower curve towards right. This flow of bacteria from the left chamber to the right chamber will continue until there's a higher chance for bacteria to enter the channel from the right and trigger a current reversal. At this point $N_1$ will move on the lower curve, $N_2$ on the upper curve and a reversed cycle will occur. When the total number of particle is equal to $N_0$, $N_1$ and $N_2$ will start from the middle of the graph (Fig. (\ref{fig:fig5}) panel c) and the two rates $\alpha$ will only become equal when one structure is full and the other is empty and both values of alpha vanish. At this point a current reversal can only occur by an unbiased random fluctuation and the process is reversed with no defined period. %This simple picture also captures the fact that the amplitude of oscillations increases when $N$ is increased (see  \cite{supp}). % \textcolor{blue}{The red curve in the inset of Fig. 3c, is the amplitude of the
%oscillations predicted by the model compared with the numerical data (blue symbols).}
%As a consequence of this, if we assume that the magnitude of channel current weakly depends on $N$, we can also deduce the observed period increase for larger $N$.
This simple picture quantitatively describes the observed increase of
oscillation amplitudes with increasing $\phi$ (see inset in Fig.3c).

\hspace{1cm}

%%%%%%%%  TWO CHAMBERS  %%%%%%%%%%%%%%%%%%%%%%%%%%%%%%%%%%%%%%%

\begin{figure}[!t]
\centering
\includegraphics[width=.5\textwidth]{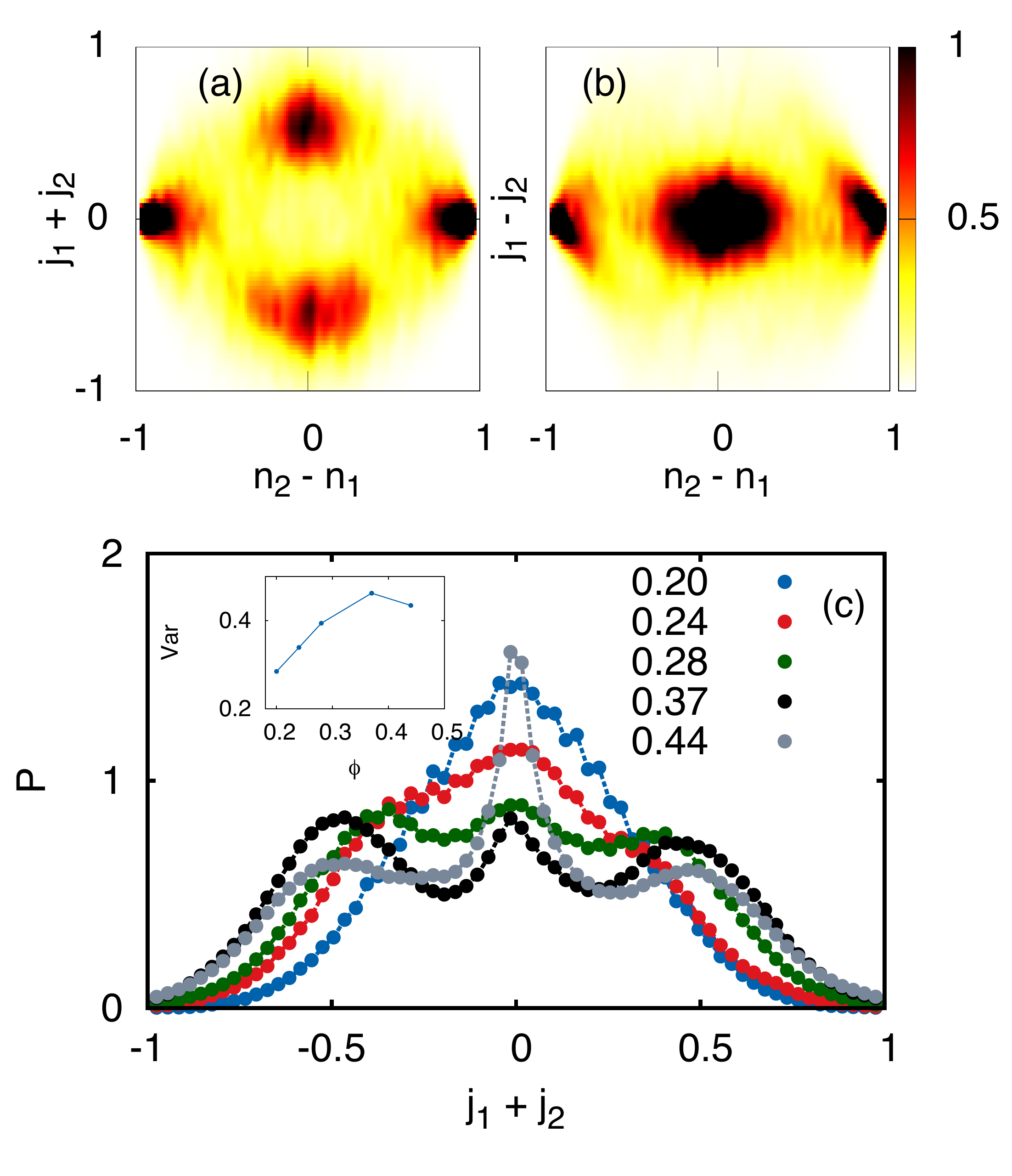}
\caption{
Double channel.
Top panels: the join probabilities $P(j_1\!+\!j_2,\delta n)$ (panel (a)) and $P(j_1\!-\!j_2,\delta n)$ (panel (b)) for $\phi=0.44$. 
Panel (c): probability distribution function $P(j_1\!+\!j_2)$
increasing the number of swimmers for $L=50\ell$ and $\sigma=\ell/2$. Inset:
variance of the distribution $P(j_1\!+\!j_2)$ as a function of $\phi$ for $L=50\ell$ and $\sigma=\ell/2$.
}
\label{fig:rho}       % Give a unique label
\end{figure}

We now consider the case in which two channels connect the chambers.
Performing  numerical simulations at different particles area fractions $\phi$,
from 0.20 to 0.58 \cite{phi},
we observe two dominant behaviors -- see 
snapshots  in panel (b) of Fig. (\ref{fig:snap}) and
the movies in \cite{supp}.
The first one is the oscillatory pumping of swimmers between the two chambers
-- top of panel (b) in Fig. (\ref{fig:snap}) -- in analogy with what observed in
the presence of one channel.
Again, differently from the equilibrium case, 
the fraction of particles in the containers is affected by strong fluctuations
that favour flows that alternately empty and fill the chambers.
The presence of the second channel has the effect of reinforce the flows,
resulting in higher peak frequencies (shorter oscillation times)
with respect to the one-channel case.
The second new behavior, forbidden by construction when only one channel connects the chambers,
is a circular flow of swimmers in the system.
In such a case  the currents $j_1$ and $j_2$ in the two channels have the same orientation 
and the number of particles $n_1$ and $n_2$ inside the chambers turn out to be nearly constant.
This circulating flow persists until a fluctuation in the number of 
right and left oriented particles in a channel gives rise to an inversion of the flux,
thus resulting in two channel flows pointing towards the same chamber falling back into
the previous situation of oscillatory pumping  behavior.
This kind of intermittent behavior between the two phases 
is made clear 
by looking at the time evolution of  $n_i$ and $j_{i}$
-- see as example the reported case in Fig. \ref{fig:circular}, panel (c) and (d).
When $j_1$ and $j_2$ have the same sign,  $j_1 \cdot j_2 >0$,
 a circular flow sets in (grey area in the figure),
with a nearly constant population level in the two containers.
In the case $j_1 \cdot j_2 <0$ swimmers flow into the same box, increasing its population level
until a reverse flux pumps out particles from it and fills the other container (white area in the figure).
The existence of circulating flows and oscillatory pumping 
is also evident by looking at the joint probability densities
$P(j_1\!+\! j_2,\delta n)$ and 
$P(j_1\!-\! j_2,\delta n)$.
Circulating flows correspond to the two spots of $P(j_1\!+\!j_2,\delta n)$ 
at $\delta n \sim 0$ and $j_1+j_2 \neq 0$ (Fig. (\ref{fig:rho})a)
and to the single spot of $P(j_1\!-\!j_2,\delta n)$ at the origin (Fig. (\ref{fig:rho})b).
The oscillatory behavior produces, instead, accumulation of particles into 
the chambers, producing the spots at $\delta n \sim \pm 1$.

The role of swimmers density is analyzed by studying the $\phi$-dependence of the probability density
 $P(j_1\!+\!j_2)$.
We observe that $P(j_1\!+\!j_2)$ changes its shape by increasing $\phi$, developing a three
peaks structure at high densities, $\phi>0.24$ -- Fig. (\ref{fig:rho}), panel c.
The two peaks at $j_1\!+\!j_2\neq 0$ are due to the circulating flows. 
The peak in zero corresponds to the limiting situations of maximum density unbalance produced by the alternating pumping state (the spots at $\delta n\sim\pm1$ and $j_1\!+\!j_2\sim0$ in panel (b) Fig. (\ref{fig:rho})).  
This central component increases with particle density producing a non monotonic behaviour 
in the variance of $P(j_1\!+\!j_2)$ as a function of $\phi$ (see the inset in 
 Fig. (\ref{fig:rho})c). 
Changing the chamber size at fixed density does not have relevant effects on the reported behavior \cite{supp}.
%
%
%
%
%%%%%%%%%%%%%%%%%%%%%%%%%%%%%%%%%%%%%%%%%%%%%%%%%%%%%%%%%%%5
%

{\it Conclusions.}---
We have shown that confining active particles in micro-chambers connected by thin channels gives rise to
self-sustained density oscillations, alternately filling the two containers. 
When adding a second channel, this oscillatory behaviour is still present  together with circulating flows.
The basic ingredients of such an effect are a density dependent motility combined with the narrowness of the channels, which only permits single file dynamics and inhibits cell reorientation.
The reported numerical findings suggest a possible route to generate self-sustained oscillations in active matter.
The proposed mechanism could be investigated experimentally since  bacteria such as {\it E. coli} and {\it B. subtilis} remain motile even when constrained  in  thin channels with a size that is comparable to their diameter \cite{Mannik09,Biondi98}.

\vspace{1cm}
We acknowledge support from MIUR-FIRB project
RBFR08WDBE. RDL acknowledges funding from the
European Research Council under the European Unionâs
Seventh Framework Programme (FP7/2007-2013)/ERC grant
agreement No. 307940.

%%%%%%%%%%%%%%%%%%%%%%%%%%%%%%%%%%%%%%%%

%%%%%%%%%%%%%%%%%%%%%%%%%%%%%%%%%%%%%%%%
%\clearpage


\begin{thebibliography}{99}
%
%
%
\bibitem{sync} A. Pikovsky, M. Rosenblum, and J. Kurths, 
{\itshape Synchronization - A Universal Concept in Nonlinear Sciences}
(Cambridge University Press, Cambridge, England, 2001).
\bibitem{Berg04} H. C. Berg, {\itshape E. Coli In Motion} (Springer, New York, 2004).
%\bibitem{Vicsek95} T. Vicsek, {\itshape et al.}, Phys. Rev. Lett. {\bf{75}}, 1226 (1995).
\bibitem{Marchetti13} M. C. Marchetti  {\itshape et al.}, Rev. Mod. Phys. {\bf{85}}, 1143 (2013).
\bibitem{Angelani09} L. Angelani, R. Di Leonardo and G. Ruocco, Phys. Rev. Lett. {\bf{102}}, 048104 (2009).
\bibitem{Angelani11c}L. Angelani and R. Di Leonardo, Comp. Phys. Commun. {\bf{182}}, 1970 (2011);
New J. Phys. 12, 113017 (2010).
\bibitem{DiLeonardo10} R. Di Leonardo, {\itshape et al.}, Proc. Natl. Acad. Sci. {\bf{107}}, 9541 (2010).
\bibitem{Sokolov10} A. Sokolov, {\itshape et al.}, Proc. Natl. Acad. Sci. U.S.A. {\bf{107}}, 969 (2010).  
\bibitem{Koumakis13} N. Koumakis, {\itshape et al.}, Nature Communications, {\bf{4}}, 2588 (2013).

\bibitem{Narayan07} V. Narayan, S. Ramaswamy, and N. Menon, Science {\bf{317}}, 105 (2007).
\bibitem{Toner} J. Toner, and Y. Tu, Phys. Rev. Lett. {\bf{75}}, 4326 (1995).  Phys. Rev. E {\bf{58}}, 4828 (1998).
J. Toner, Y. Tu, and S. Ramaswamy, Ann. Phys. {\bf{318}}, 170 (2005).
\bibitem{Ramaswamy03} S. Ramaswamy, R. A. Simha, and J. Toner, Europhys. Lett. {\bf{62}}, 196 (2003).

\bibitem{Sumino12} Y. Sumino {\itshape et al.} Nature {\bf{483}}, 448 (2012).
\bibitem{Schaller11} V. Schaller, {\itshape et al.}, Proc. Natl. Acad. Sci. U.S.A. {\bf{108}}, 19099 (2011). 
\bibitem{Zhou14} S. Zhou, {\itshape et al.}, Proc. Natl. Acad. Sci. U.S.A. {\bf{111}}, 1265 (2014). 
\bibitem{Bricard13}A. Bricard, J.-B. Caussin, N. Desreumaux, O. Dauchot, and D. Bartolo, Nature {\bf{503}}, 95 (2013).


%\bibitem{Galajda07} P. Galajda {\itshape et al.}, J. Bacteriol. {\bf{189}}, 1033 (2007).
\bibitem{Sanchez12} T. Sanchez, {\itshape et al.}, Nature {\bf{491}}, 431 (2012).
\bibitem{wioland} H. Wioland, F.G. Woodhouse, J. Dunkel, J.O. Kessler, and R.E. Goldstein
Phys. Rev. Lett. {\bf 110}, 268102 (2013).
%
%
%\bibitem{Angelani11} L Angelani, {\itshape et al.}, Phys. Rev. Lett. {\bf{107}}, 138302 (2011).
%\bibitem{Wu00}  X. L. Wu and A. Libchaber, Phys. Rev. Lett. {\bf{84}}, 3017 (2000).

%\bibitem{Hol14} F. J. H. Hol, and C. Dekker, Science {\bf{346}}, 1251821 (2014). 
%
%\bibitem{Wioland13} H. Wioland {\itshape et al.} Phys. Rev. Lett. {\bf{110}}, 268102 (2013).
\bibitem{Galajda14} O. Sipos, K. Nagy, and P. Galajda, Chem. Biochem. Eng. Q. {\bf{28}}, 233 (2014).
\bibitem{Jelic12} A. Jeli\'c {\itshape et al.} Eur. Phys. Lett. {\bf{98}}, 40009 (2012).
\bibitem{Illien13} P. Illien {\itshape et al.} Phys. Rev. Lett. {\bf{111}}, 038102 (2013).
\bibitem{Gorissen12} M. Gorissen {\itshape et al.} Phys. Rev. Lett. {\bf{109}}, 170601 (2012).
\bibitem{Ner13} I. Neri {\itshape et al.} Phys. Rev. Lett.  {\bf{110}}, 098102 (2013).
\bibitem{Locatelli14} E. Locatelli, M. Pierno, F. Baldovin, and E. Orlandini,  Phys. Rev. E {\bf{91}}, 022109 (2015).
\bibitem{Wei00} Q. H. Wei, C. Bechinger, and  P. Leiderer, Science {\bf{287}}, 625-627 (2000).


\bibitem{Reichhardt08} M. B. Wan, C.J. O. Reichhardt, Z. Nussinov, and C. Reichhardt, Phys. Rev. Lett.  {\bf{101}}, 018102 (2008).

\bibitem{Schnitzer93} M. J. Schnitzer, Phys. Rev. E {\bf{48}}, 2553 (1993).
\bibitem{Cates12} M. E. Cates, Rep. Prog. Phys. {\bf{75}}, 042601, (2012).
\bibitem{Tailleur08} J. Tailleur, and M. E. Cates, Phys. Rev. Lett. {\bf{100}}, 218103 (2008).
\bibitem{Angelani11b} L. Angelani, A. Costanzo and R. Di Leonardo, EPL {\bf{96}}, 68002 (2011).



%\bibitem{Paoluzzi}M. Paoluzzi, R. Di Leonardo, and L. Angelani, J. Phys.: Condens. Matter {\bf{25}} 415102 (2013). 
\bibitem{Paoluzzi}M. Paoluzzi, R. Di Leonardo, and L. Angelani, J. Phys.: Condens. Matter {\bf{25}} 415102 (2013); J. Phys.: Condens. Matter {\bf{26}} 375101 (2014). 


%\bibitem{Kim05} S. Kim and S. Karrila, {\itshape Microhydrodynamics} (Dover, New York, 2005).
%\bibitem{Dey12} S. Dey, D. Das, and R. Rajesh, Phys. Rev. Lett. {\bf{108}} 238001 (2012).
\bibitem{supp} Supplemental material, which includes Refs. \cite{Kim05} and \cite{Angelani13}.
\bibitem{Kim05} S. Kim and S. Karrila, {\itshape Microhydrodynamics} (Dover, New York, 2005).
\bibitem{Angelani13} L. Angelani, EPL 102, 20004 (2013).
\bibitem{phi} The area fraction $\phi$ is calculated as $\!N A_s / A$  where 
$A_s\!=\!\pi(l/4)^2\!+\!l^2/4$ is the cross section of a single spherocylindrical swimmer and 
$A\!=\!2 \pi R^2\!+\!\sigma L$ is the total available area
($A\!=\!2 \pi R^2\!+\!2\sigma L$ in the case of the double channel).
\bibitem{Biondi98} S. A. Biondi, {\it et al.} AICHE J {\bf{44}}, 1923 (1998).
\bibitem{Mannik09} J. Mannik {\itshape et al.}, Proc. Natl. Acad. Sci. U.S.A. {\bf{106}}, 1468 (2009).

%\bibitem{KL51} S. Kullback, and R. A. Leibler, Ann. Math. Stat {\bf{22}}, 79 (1951).
%\bibitem{Henkes11} S. Henkes, Y. Fily, and M. C. Marchetti, Phys. Rev. E {\bf{84}}, 040301 (R) (2011).
%
%\bibitem{Paoluzzi14} M. Paoluzzi, R. Di Leonardo, and L. Angelani, arXiv:1401.6944 (2014).
%%
%
%\bibitem{Palacci10} J. Palacci, {\itshape et al.}, Phys. Rev. Lett. {\bf{105}}, 088304 (2010).
%%
%%
%\bibitem{Finlayson69} B. A. Finlayson, L. E. Scriven, Proc. R. Soc. Lond. Ser. A {\bf{310}}, 183 (1969).
%\bibitem{Berg72} H. C. Berg, D. A. Brown, Nature (London) {\bf{239}}, 500 (1972).
%%
%\bibitem{Cavagna10} A. Cavagna, {\itshape et al.}, Proc. Natl. Acad. Sci. U.S.A. {\bf{107}}, 11 865 (2010).
%\bibitem{Makris06} N. C. Makris, {\itshape et al.}, Science {\bf{311}} 660 (2006).
%\bibitem{Silverberg13} J. L. Silverberg, {\itshape et al.}, Phys. Rev. Lett. {\bf{110}}, 228701 (2013).
%\bibitem{Dombrowski04} C. Dombrowski, {\itshape et al.}, Phys. Rev. Lett. {\bf{93}}, 098103 (2004).
%
%\bibitem{Buttinoni13} I. Buttinoni, {\itshape et al.}, Phys. Rev. Lett. {\bf{110}}, 238301 (2013).
%%
%
%%
%%
%%
%\bibitem{Berg72} H. C. Berg, D. A. Brown, Nature (London) {\bf{239}}, 500 (1972).
%\bibitem{Grossart01} H. -P. Grossart, L. Riemann, and F. Azam, Aquatic Microbial Ecology {\bf{25}}, 247 (2001).
%\bibitem{Angelani10} L. Angelani and R. Di Leonardo, New Journal of Physics {\bf{12}}, 113017 (2010).
%\bibitem{Farrell12} F. D. C. Farrell, M. C. Marchetti, D. Marenduzzo, and J. Tailleur, Phys. Rev. Lett. {\bf{108}}, 248101 (2012).
%\bibitem{Cates10} M. E. Cates, D. Marenduzzo, I. Paganobarraga, and J. Tailleur, Proc. Natl. Acad. Sci. U. S. A. {\bf{107}}, 11715 (2010).
%\bibitem{Purcell77} E. M. Purcell, Am. J. Phys. {\bf{45}}, 3 (1977).
%\bibitem{Kim05} S. Kim and S. Karrila, {\itshape Microhydrodynamics} (Dover, New York, 2005).
%\bibitem{Press92} W. H. Press, W. T. Wetterling, S. A. Teukolsky, B. P. Flannery, Numerical Recipes in C, Cambridge University Press, Cambridge, (1992).
 %{\bf{4}} (1971) 1071.
%
%
%
\end{thebibliography}
\end{document}